# Single Layer Behavior and Its Breakdown in Twisted Graphene Layers

A. Luican[1], Guohong Li[1], A. Reina[2], J. Kong[3], R. R. Nair[4], K. S. Novoselov[4], A. K. Geim[4,5], E.Y. Andrei[1]

[1]Department of Physics and Astronomy, Rutgers University, Piscataway, New Jersey 08855,USA

[2]Department of Materials Science and Engineering, MIT, Cambridge, Massachusetts 02139,USA

[3]Department of Electrical Engineering and Computer Science, MIT, Cambridge, Massachusetts 02139, USA

[4]School of Physics and Astronomy, University of Manchester, M13 9PL, Manchester, UK

[5]Manchester Centre for Mesoscience and Nanotechnology, University of Manchester, M13 9PL, Manchester, UK

We report high magnetic field scanning tunneling microscopy and Landau level spectroscopy of twisted graphene layers grown by chemical vapor deposition. For twist angles exceeding ~ 3° the low energy carriers exhibit Landau level spectra characteristic of massless Dirac fermions. Above 20° the layers effectively decouple and the electronic properties are indistinguishable from those in single layer graphene, while for smaller angles we observe a slow-down of the carrier velocity which is strongly angle dependent. At the smallest angles the spectra are dominated by twist induced Van Hove singularities and the Dirac fermions eventually become localized. An unexpected electron-hole asymmetry is observed which is substantially larger than the asymmetry in either single or untwisted bilayer graphene.

One of the remarkable aspects of the 2d relativistic quasiparticles (massless Dirac fermions) in graphene[1, 2] is their chiral symmetry [3] which is revealed in the presence of a magnetic field, $B$, through the appearance of a unique Landau level (LL) at an energy that is pinned to the Dirac point. The other LLs exhibit the distinct signature of the relativistic quasiparticles through a square-root dependence on field and level index, $n$. When graphene layers stack together, as is often the case in CVD grown [4-6] and epitaxial graphene [7], interlayer coupling typically destroys the relativistic quasiparticles and produces new types of excitations with LL sequences that reflect the number of layers and degree of coupling [8-11]. Thus it was quite surprising that in multi-layer graphene grown on SiC, ARPES and STM measurements showed spectra corresponding to massless Dirac fermions in single graphene layers [12-14]. The reason for their survival in this multi-layer system, initially attributed to the decoupling of twisted layers [10, 15] is intensely debated [16-22]. On the theoretical side, *ab initio* calculations predicted that all misoriented graphene layers are effectively decoupled [21] while tight-binding calculations showed that interlayer coupling still plays an important role in renormalizing the Fermi velocity of the massless Dirac fermions by an amount which depends on the twist angle[10, 16, 22]. On the experimental side, transport experiments [7, 18] provided evidence for a reduced Fermi velocity in twisted layers while Raman studies [19, 20] are still unsettled. Scanning tunneling microscopy and spectroscopy (STM/STS) are the tools of choice to address this issue because they can simultaneously measure the local twist angle, the Fermi velocity and the degree of interlayer coupling. The first is obtained from the period of the moiré pattern imaged by STM [23] while the last two are accessed through LL spectroscopy measured by STS [24].

In this Letter, we combine high magnetic field STM topography with LL spectroscopy to study the effect of twisting on the band structure. For twist angles above $20^0$ the electronic properties of the twisted layers are indistinguishable from single layer graphene. At smaller twist angles we observe a strong angle dependent downward renormalization of the Fermi velocity in quantitative agreement with theoretical predictions [10]. We further find that the twist causes a rather large electron-hole asymmetry not considered so far theoretically. At the smallest twist angle ~1.16°, we find that the massless Dirac fermion picture breaks down and Van Hove singularities dominate the electronic spectrum.

Experiments were conducted in a home built, low temperature (T=4.4K for this work) high field STM using mechanically cut Pt-Ir tips. The tunneling conductance, dI/dV, was measured by lock-in detection with 340 Hz bias voltage modulation ($2mV_{rms}$ otherwise specified). The magnetic field was applied perpendicular to the sample surface with a superconducting magnet working in persistent mode. Samples were large area films of few-layer graphene grown via ambient pressure chemical vapor deposition (CVD) on polycrystalline Ni films [4]. The films are continuous and can be patterned lithographically. A 250nm layer of PMMA was spin-coated on graphene to provide mechanical support, after which the graphene layers with PMMA on top were released from the Ni film by etching in 1M $FeCl_3$ solution. After the release, the graphene layer covered with PMMA is transferred to a TEM grid. We used acetone to dissolve the PMMA and thus release the graphene on the gold grid. The samples were finally dried in a critical point dryer to prevent the membrane from rupturing due to surface tension. STM experiments were performed on both suspended and non-suspended regions of the sample. The suspended areas were too large (~30μm) to be mechanically stable in the presence of an approaching STM tip. The data reported here were taken on non-suspended parts of the sample.

TEM studies indicate that most CVD graphene films contain twisted layers with twist angles larger than 10° [4]. Fig.1(a) shows an STM image of a typical twisted area. We use Landau level spectroscopy [24] to characterize the electronic states of the film because it provides a sensitive probe of the quasiparticle excitation spectrum allowing to distinguish between massive and massless Dirac fermions and to determine the degree of interlayer coupling [11, 25, 26]. The LL spectrum of massless Dirac fermions follows the sequence,

$$E_n = E_D + \text{sgn}(n)\sqrt{2e\hbar v_F^2 |n| B}, \quad n = 0, \pm 1, \pm 2.. \quad (1)$$

where $E_D$ is the energy at the Dirac point, $-e$ is the electron charge, $\hbar$ the Planck constant over $2\pi$ and $v_F$ the Fermi velocity.

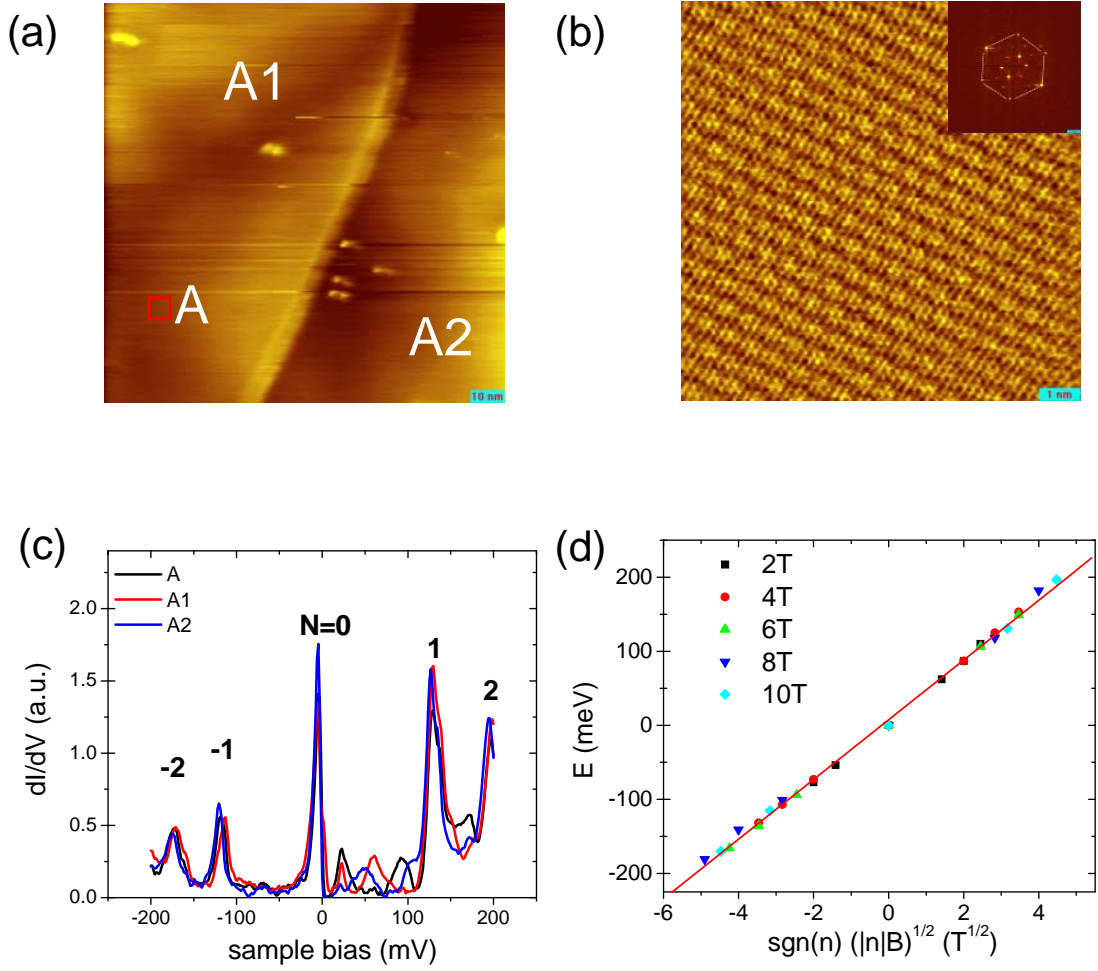

FIG. 1. Combined STM and LL spectroscopy of twisted graphene layers. (a) Large area topography taken with tunneling junction setting of 300mV and 25pA (same in (c)). Scale bar: 10nm. (b) Zoom-in topography of region A (red square) shows a moiré pattern corresponding to a twist angle of $\theta=(21.8\pm1.7)°$ Inset: Fourier transform of the main panel showing the contributions from the atomic lattice (outer hexagon) and from the superstructure (inner hexagon).(c) Tunneling spectra taken at different regions marked in (a) show similar Landau levels of massless Dirac fermions in a magnetic field of 10 T. Scale bar: 1nm. (d) Scaling of LL's in region A. Symbols are the LL's in different magnetic fields. Solid line is a linear fit according to Eq.(1). $n$ the is level index with $n>0$ corresponding to electrons and $n<0$ to holes.

The LL peaks are clearly seen in the tunneling spectra shown in Fig.1(c) for three regions of the sample in a field of 10T. We note that this same sequence appears across the entire twisted area. The field dependence of the LL energies in region A of Fig.1(a), plotted in Fig.1(d) against the reduced parameter $(|n|B)^{1/2}$, reveals the characteristic scaling in Eq.(1), expected for massless Dirac fermions. The Fermi velocity, obtained from the slope of the linear fit $v_F$=(1.10±0.01)×$10^6$m/s, is consistent with that in graphene layers grown on the C-face of SiC [14].

In region A, high resolution topography (Fig.1b), shows a superstructure, whose Fourier transform (inset) reveals two sets of peaks arranged in concentric hexagons which correspond to the atomic lattice and to the superstructure for the outer and inner hexagons respectively. The relative rotation angle in k-space between the two hexagons is ~20°. Such a superstructure, or moiré pattern, forms as a result of the twist between the layers. The twist angle $\theta$ in real space is related to the rotation angle $\varphi$ in k-space by $\varphi$=30°-($\theta$/2). For $\varphi$ ~ 20°, $\theta$ ~ 20°. A better estimate of the twist angle can be obtained from the period of the moiré pattern $L$,

$$L = a /(2 \sin(\theta / 2)) \qquad (2)$$

where $a$ = 0.246nm the lattice constant of graphene. $L$= 0.65 ± 0.05nm in Fig.1(b) gives $\theta$ = (21.8±1.7)°, consistent with the above estimate in k-space. Eq.(2) is valid for $\theta$ < 30°; a twist angle of $\theta$ = 38.2° is expected to produce a pattern with the same period as $\theta$ = 21.8°, but the spectrum of this rotated structure is also expected to have a gap ~ 10meV near the Dirac point, which is absent for $\theta$ = 21.8° . In our case, the single un-split peak observed at the Dirac point, Fig.1(b), is therefore consistent with $\theta$ = 21.8°. In the following we focus on the main features of the tunneling spectra and consider only angles $\theta$ < 30° because $\theta$'= (60°- $\theta$) and $\theta$ produce similar moiré patterns and have the same effect on the velocity renormalization [17, 21]. The small peaks in between LL observed in Fig.1(c) could be related to fine electronic structure in twisted layers [17, 21].

Landau levels similar to those in Fig.1(c) were rather easy to observe in most regions of the film, but resolving the corresponding moiré pattern especially for a small period such as that in Fig.1(b) is more challenging and requires a sharper STM tip. A resolved moiré pattern with a larger period of $L$=4.0 ±0.2 nm, corresponding to $\theta$=3.5±0.3°, is shown in Fig.2(b). We measured the field dependence of the LL spectra in this region (region B in Fig 2a) and its adjacent region C and repeating the procedure described above we find that in both regions the LL spectra are well described by the massless Dirac Fermion sequence of Eq.(1) with $v_F$ = 0.87×$10^6$m/s for region B and $v_F$= 1.10×$10^6$m/s for region C .

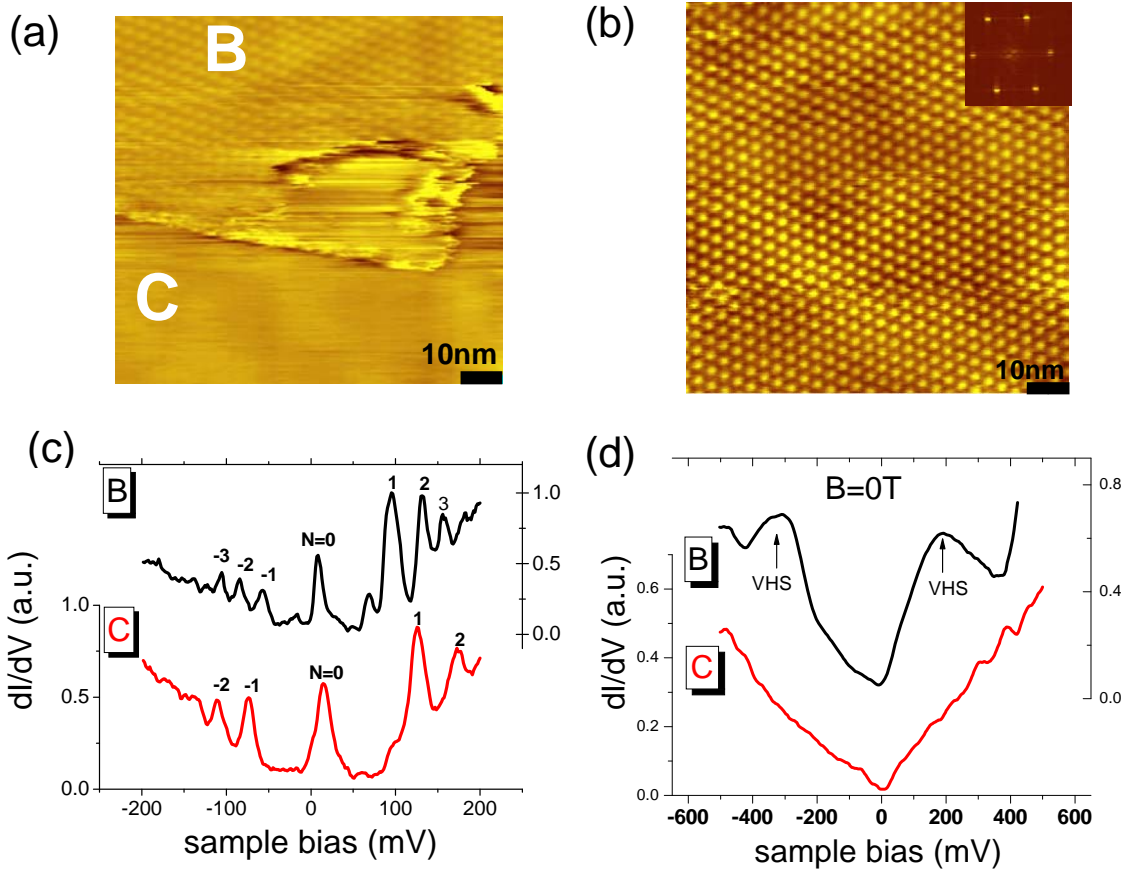

FIG. 2 (a) Large area topography showing the boundary between a region with a moiré pattern (B) and region C, which is featureless. (b) Topography in the center of the area with the moiré pattern. The pattern period, L=(4.0±0.2)nm, corresponds to a twist-angle of θ=(3.5±0.3)° . Scale bar: 1nm. Inset: Fourier transform of the main panel showing the contributions from the atomic lattice (outer hexagon) and from the superstructure (inner hexagon). (c) High field (6T) tunneling spectra taken at positions B and C in (a). (d) Zero field tunneling spectra taken at positions B and C in (a). Tunneling junction setting: 300mV and 20pA for (a) and (b); 500mV and 20pA for (d).

Why is it that the low energy physics of the quasiparticles in twisted graphene layers can resemble that in a single layer? To answer this question we consider the (K and K') corners of the hexagonal Brillouin zone of single layer graphene where the Dirac cones reside [1]. When two layers are superposed with a relative twist, the corresponding Brillouin zones also rotate with respect to each other so that the Dirac cones for the two layers separate (Fig.3 (a) inset) at low energy by an amount which increases with angle: $\Delta K=2K\sin(\theta/2)$, where $K=4\pi/3$. The two displaced cones cross at a higher energy and, in the presence of interlayer coupling, merge into a saddle point [10, 27]. For large twist angles, when the crossing energy is sufficiently far from the Dirac point, the low energy part of the Dirac cones and the corresponding physics should be indistinguishable from that of a single layer. As the twist angle decreases, the Dirac cones are modified by the proximity to the saddle point and, for $\theta > 3^0$, the excitation spectrum can still be described by massless Dirac fermions but with a renormalized Fermi velocity given by[10]:

$$\frac{v_F(\theta)}{v_F^0} = 1 - 9\left(\frac{t_\perp^\theta}{\hbar v_F^0 \Delta K}\right)^2 \qquad (3)$$

where $t_\perp^\theta \approx 0.4 t_\perp$ and $t_\perp$ are the interlayer coupling for Bernal stacking. As shown in Fig.3(a), the velocity renormalization measured in our experiment is in good agreement with the predictions of Eq.(3).

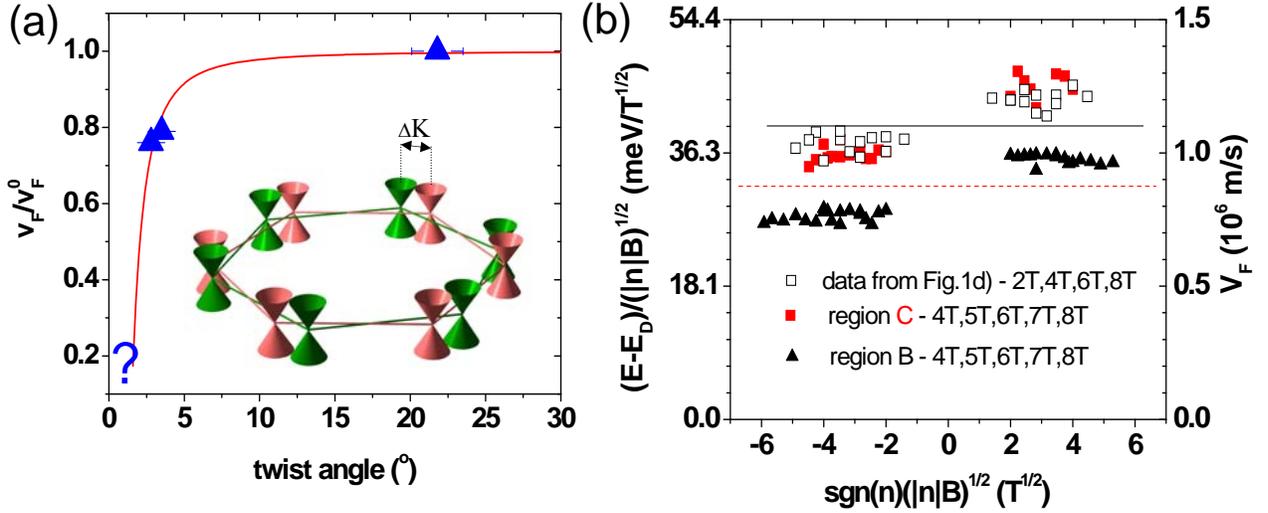

FIG. 3. Fermi velocity renormalization. (a) Angle dependence of the renormalization. Line is theoretical prediction according to Eq.(3). Triangles are experimental data. Question mark at ~1.16° corresponds to localized states discussed in Fig.4. Inset, Dirac cones of twisted layers. The twist-induced separation between the Dirac cones in the two layers, ΔK, is controlled by the angle. (b) Electron hole asymmetry of Fermi velocity in different sample regions is independent of field or level index. Symbols are Fermi velocities obtained by Eq.(4) and solid lines are overall fitting according to Eq.(1).

We now compare the results to previous STM/STM studies on graphene layers on SiC [14] which reported that the LL sequences were independent of the measured moiré pattern periods for a wide range of twist angles down to ~1.4°. This appeared to be in direct contradiction to the theoretical predictions [10, 17, 21]. An important clue to understanding these results can be found in the unusual presence of the same continuous atomic honeycomb structure across the entire superstructure. This is in sharp contrast to moiré patterns generated by two rotated layers where one sees a close correlation between the super-pattern and the atomic structure which changes continuously from triangular to honeycomb across each period of the pattern [23, 27]. In fact, as was shown later [28] the large period superstructure reported in ref. [14] was actually produced by deeper layers below the surface, serving as a background, whereas the twist angle between the top layer and the layer below it was large. With this interpretation, the data reported in [14] is consistent both with the theoretical models and with the results reported here.

We next consider the states close to the saddle points. Such saddle points cause a divergence in the density of states, also known as van Hove singularities. In Fig.2(d), we compare the zero field tunneling spectra over a large sample bias range for regions B and C of Fig.2(a). The spectrum in region B shows two peaks separated [27] by ~200 meV in good agreement with the position of the expected van Hove singularities for the observed twist angle of ~ 3.5°. In contrast for region C the plain V-shape spectrum is consistent with a large twist angle whose corresponding moiré pattern is not within experimental resolution.

It is important to emphasize that the twist-induced velocity renormalization in coupled graphene layers is different from that due to electron-phonon interactions observed in single layer graphene supported on graphite [24]. Although both phenomena produce a downward velocity renormalization, the associated spectral features and the underlying physics are completely different. The twist-induced slow-down described here, produces two pronounced peaks in the zero field density of states separated by an energy that increases monotonically with twist-angle. By contrast, slow-down due electron-phonon interactions produces kinks that are separated from the Dirac point by an energy corresponding to the $A_1'$ phonon. These kinks reflect strong electron coupling due to the Kohn anomaly of this phonon, which modifies the slope of the Dirac cone and reduces the Fermi velocity. Interestingly according to *ab initio* calculations [29] the electron-phonon interaction is strongly suppressed in multi-layer graphene in the presence of coupling between layers. Indeed this is consistent with the observation of a reduced $v_F \sim 0.79 \times 10^6$ m/s [24] in single layer graphene compared to $1.07 \times 10^6$ m/s in coupled multilayers [11, 14]. We note that for the twisted layers discussed in Fig.2(c), $v_F$ is almost identical to that in multi-layers with Bernal stacking, suggesting that electron-phonon coupling via $A_1'$ is also suppressed in twisted layers.

Next we take a closer look at the Fermi velocity in twisted layers. Eq.(1) can be rewritten as

$$v_F = (E_n - E_D)/\text{sgn}(n)\sqrt{2e\hbar |n| B}, \quad n = 0, \pm 1, \pm 2.. \quad (4)$$

so that, instead of an overall fitting to Eq.(1), we plot $v_F$ for electrons an holes separately as in Fig.3(b). Now surprisingly $v_F$ is not the same for electrons and holes and the electron carriers are systematically faster than the holes. Thus in regions A (($\theta \sim 21.8°$) and C where the average velocity is $v_F = 1.10 \times 10^6$m we find an electron hole asymmetry with $v_F = 1.20 \times 10^6$m/s for electrons and $v_F = 1.02 \times 10^6$m/s for holes. In region B ($\theta \sim 3.5°$) where the average velocity is $v_F = 0.87 \times 10^6$m we measure $v_F = 1.00 \times 10^6$m/s for electrons and $v_F = 0.76 \times 10^6$m/s for holes. The electron-hole asymmetry is larger at the smaller angles, ±14% for 3.5°, compared to ±8% for 21.8°. For comparison we note that the asymmetry is less than ±1% in single layer graphene suspended over a graphite surface [24] and ±2.5% for single layer graphene on $SiO_2$. The latter was explained by a large nearest-neighbor overlap integral [30]. The origin of the significantly larger asymmetry in the twisted layers has not been studied. One likely cause for this asymmetry is the enhanced next-nearest neighbor hopping enabled by the twist, but further work is needed to understand this phenomenon.

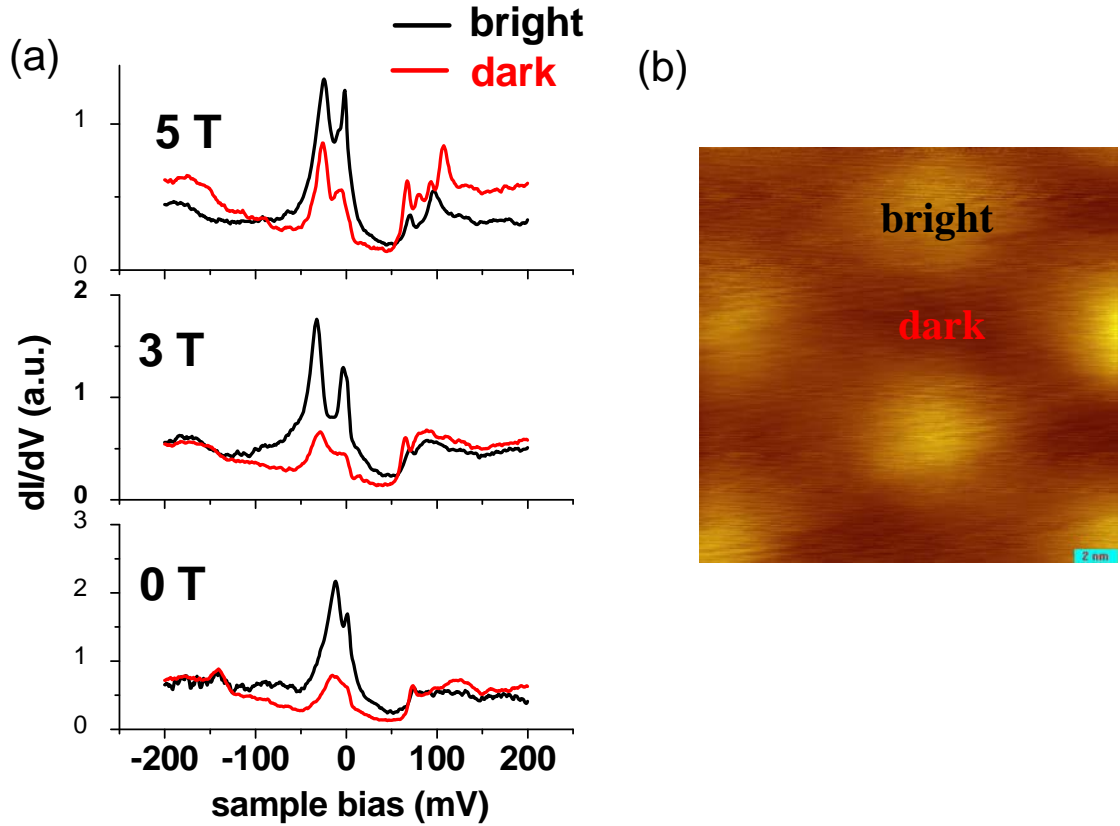

FIG. 4. (a) Tunneling spectra and their field dependence for a very small twist angle ~1.16°, corresponding to the question mark in Fig. 3(a). The large difference between the spectra in the dark and bright regions diminishes with increasing field. No Landau levels of massless Dirac fermions were observed here. (b) STM image of the moiré pattern with a period of 12 nm. Scale bar: 2nm. Tunneling junction setting: 200mV and 20pA.

When the twist angle is very small, the velocity renormalization picture no longer applies because the Van Hove singularities start dominating the spectrum and the Dirac cone approximation breaks down even at the lowest energies [27] as is clearly seen in Fig.4(a). In this regime the spectra are strongly spatially modulated in registry with the moiré pattern, indicating that the carriers become localized in a charge density wave (CDW) [27]. As seen in Fig.4(a), the contrast between spectra in the dark and bright regions diminishes with increasing magnetic field suggesting a competition between twist-induced localization and cyclotron motion. This is consistent with the fact that the magnetic length, $l_B = \sqrt{e/(\hbar B)} = 11.5 nm$, becomes comparable to the period of the moiré pattern, 12nm, at 5T and provides another clue to the physics of the twisted layers.

In summary, by using STM together with LL spectroscopy we demonstrated that the low energy electronic properties of twisted graphene layers are controlled by masssless Dirac fermions whose Fermi velocity is renormalized by the twist. This picture breaks down at the smallest twist angles where the spectrum is taken over by twist-induced Van Hove singularities which favor the formation of a CDW. Finally, the observation in CVD grown graphene of remarkably clean LL sequences corresponding to massless Dirac fermions is attests to the high quality and potential of this material for electronic applications.

E.Y.A. acknowledges DOE support under DE-FG02-99ER45742 and partial NSF support under NSF-DMR-0906711 and. A.R. and J.K. acknowledge partial support of NSF DMR 0845358 and support under ONR MURI N00014-09-1-1063. The Manchester contributors acknowledge financial support from EPSRC, Office of Naval Research and the Royal Society.

[1] A. H. Castro Neto *et al.*, Reviews of Modern Physics **81**, 109 (2009).
[2] D. S. L. Abergel *et al.*, Advances in Physics **59**, 261
[3] G. W. Semenoff, Physical Review Letters **53**, 2449 (1984).
[4] A. Reina *et al.*, Nano Letters **9**, 30 (2008).
[5] X. Li *et al.*, Science **324**, 1312 (2009).
[6] K. S. Kim *et al.*, Nature **457**, 706 (2009).
[7] W. A. de Heer *et al.*, Solid State Communications **143**, 92 (2007).
[8] F. Guinea, A. H. Castro Neto, and N. M. R. Peres, Physical Review B **73**, 245426 (2006).
[9] E. McCann, and V. I. Fal'ko, Physical Review Letters **96**, 086805 (2006).
[10] J. M. B. Lopes dos Santos, N. M. R. Peres, and A. H. Castro Neto, Physical Review Letters **99**, 256802 (2007).
[11] G. Li, and E. Y. Andrei, Nat Phys **3**, 623 (2007).
[12] M. L. Sadowski *et al.*, Physical Review Letters **97**, 266405 (2006).
[13] M. Sprinkle *et al.*, Physical Review Letters **103**, 226803 (2009).
[14] D. L. Miller *et al.*, Science **324**, 924 (2009).
[15] J. Hass *et al.*, Physical Review Letters **100**, 125504 (2008).
[16] G. Trambly de Laissardiere, D. Mayou, and L. Magaud, Nano Letters **10**, 804 (2010).
[17] E. J. Mele, Physical Review B **81**, 161405.
[18] H. Schmidt *et al.*, Physical Review B **81**, 121403.
[19] Z. Ni *et al.*, Physical Review B **77**, 235403 (2008).
[20] P. Poncharal *et al.*, Physical Review B **79**, 195417 (2009).
[21] S. Shallcross, S. Sharma, and O. A. Pankratov, Physical Review Letters **101**, 056803 (2008).
[22] S. Shallcross *et al.*, Physical Review B **81**, 165105 (2010).
[23] P. Wing-Tat, and D. Colm, Journal of Physics D: Applied Physics **38**, R329 (2005).

[24] G. Li, A. Luican, and E. Y. Andrei, Physical Review Letters **102**, 176804 (2009).
[25] A. Luican, G. Li, and E. Y. Andrei, Solid State Communications **149**, 1151 (2009).
[26] J. M. Pereira, F. M. Peeters, and P. Vasilopoulos, Physical Review B **76**, 115419 (2007).
[27] G. Li *et al.*, Nat Phys **6**, 109.
[28] D. L. Miller *et al.*, Physical Review B **81**, 125427.
[29] J.-A. Yan, W. Y. Ruan, and M. Y. Chou, Physical Review B **79**, 115443 (2009).
[30] R. S. Deacon *et al.*, Physical Review B **76**, 081406 (2007).